\documentclass{article}
\usepackage{amsmath}
\usepackage{amsfonts}

\usepackage{amsmath,amssymb}
\usepackage{graphicx}
\usepackage{color}

\newtheorem{theorem}{Theorem}
\newtheorem{lemma}{Lemma}

\makeatletter

\newcommand{\Rmnum}[1]{\expandafter\@slowromancap\romannumeral #1@}
\makeatother

\newcommand{\F}{\ensuremath{\mathbb F}}
\newcommand{\Z}{\ensuremath{\mathbb Z}}

\newcommand{\C}{\ensuremath{\mathbb C}}

\newcommand{\done}{\hfill $\Box$ }

\newcommand{\ls}[1]
    {\dimen0=\fontdimen6\the\font\lineskip=#1\dimen0
     \advance\lineskip.5\fontdimen5\the\font
     \advance\lineskip-\dimen0
     \lineskiplimit=0.9\lineskip
     \baselineskip=\lineskip
     \advance\baselineskip\dimen0
     \normallineskip\lineskip\normallineskiplimit\lineskiplimit
     \normalbaselineskip\baselineskip
     \ignorespaces}

\begin{document}

\bibliographystyle{abbrv}

\title{A Note on the Diagonalization of the Discrete Fourier Transform}
\author{Zilong Wang$^{*1,2}$ and Guang Gong$^2$\\
$^1$ School of Mathematical Sciences, Peking University, \\
Beijing, 100871,  P.R.CHINA\\
$^2$ Department of Electrical and Computer Engineering, University of Waterloo \\
Waterloo, Ontario N2L 3G1, CANADA \\
Email: wzlmath@gmail.com \ \ \ \ ggong@calliope.uwaterloo.ca\\
}

 \maketitle

\footnotetext[0] {The work is supported by NSERC Discovery Grant.
$^*$Zilong Wang is currently a visiting Ph.D student at the
Department of ECE in University of Waterloo from September 2008 to
August 2009. He is partially supported by NKBRPC (2004CB318000).}

\thispagestyle{plain} \setcounter{page}{1}

\begin{abstract}

Following the approach developed by S. Gurevich and R. Hadani, an
analytical formula of the canonical basis of the DFT is given for
the case $N=p$ where $p$ is a prime number and $p\equiv 1$ (mod 4).

\end{abstract}

\ls{1.5}
\section{Introduction}

The Discrete Fourier transform (DFT) has important applications in
communication systems, and can be considered as an $N$-dimensional
unitary operator $F$ acting on the Hilbert space
$\mathcal{H}=\C(\Z_N)$ by the formula
$$F[\varphi](j)=\frac{1}{\sqrt{N}}\sum_{i\in \Z_N}e^{\frac{2\pi
i}{N}ij}\varphi(i).$$

In the signal processing, the time domain and frequency domain are
transformed by the DFT. A canonical basis, in other words, an
orthonormal basis of eigenvectors for $F$ will simplify the
computation of the DFT. The main difficulty to get such a canonical
basis is that $F$ is an operator of order $4$, and it has four
distinct eigenvalues $\pm 1, \pm i$ with large multiplicity if the
dimension $N>4$. The multiplicity of these eigenvalues depends on
the value of $n$ modulo $4$, and was solved in \cite{McClellan},
although it was later shown to have been equivalent to a problem
solved by Gauss in \cite{Dickinson}. Unfortunately, no simple
analytical formula for the eigenvectors is known. The research for
finding different choices of eigenvectors, selected to satisfy
useful properties like orthogonality and to have simple forms, has
been flourished in the literature \cite{McClellan} \cite{Dickinson}
\cite{Atakishiyev} \cite{Candan} \cite{Hanna}, just listed a few
here.

A novel representation theoretic approach to the diagonalization
problem of DFT in the case when $N=p$ is an odd prime number was
introduced by Gurevich and Hadani in \cite{S.Gurevich4}. This
approach introduces the Weil representation \cite{A.Weil} of the
finite symplectic group $Sp=SL_2(\F_p)$ (will be precisely defined
in Section 2) as the fundamental object of underlying harmonic
analysis in the finite setting. More precisely, a centralizer
subgroup of the DFT operator $F$ in $U(\mathcal{H})$ (see definition
in Section 2) is effectively described by using the Weil
representation, which in this setting is a unitary representation
$\rho: SL_2(\F_p)\rightarrow U(\mathcal{H})$ and the DFT is
proportional to a single operator $\rho(w)$ where $w \in
SL_2(\F_p)$. The centralizer subgroup of $w$ is $T_w$ which is a
maximal algebraic torus (i.e., maximal commutative subgroup) in
$SL_2(\F_p)$. Then $F$ commutes with $\rho(T_w)$, and they share the
same eigenvectors.

By the above approach, a canonical basis $\Phi_p$ of eigenvectors of
the DFT and the transition matrix $\Theta_p$ from the standard basis
to $\Phi_p$ ({\em discrete oscillator transform}) for $p\equiv 1$
(mod $4$) were described by an algorithm in \cite{S.Gurevich4}.
However, this algorithm has heavy computation cost, and the
analytical formulas of eigenvectors of the Fourier matrix $F$ were
unknown.

The vectors associated to the tori share many nice properties (see
\cite{S.Gurevich}  \cite{S.Gurevich2} \cite{S.Gurevich3} for recent
applications) and a simple analytical formula for the vectors
associated to split tori was given in \cite{Wang}. Based on
\cite{S.Gurevich4} and \cite{Wang}, in this paper, an analytical
formula of the canonical basis of the DFT for the case of $p\equiv
1$ (mod $4$) is given in Theorem 3, and their respective
corresponding eigenvalues are determined in Theorem 4. Then the
discrete oscillator transform $\Theta_p$ introduced in
\cite{S.Gurevich4} can be obtained in a straightforward manner.

The rest of the paper is organized as follows. In Section 2, we
introduce the definitions of the one dimensional finite Heisenberg
and Weil representations and the approach studying the eigenvectors
of the DFT exhibited in \cite{S.Gurevich4}. Then in Section 3, we
give an analytical formula of the canonical basis of the DFT, and
determine their respective corresponding eigenvalues.

\section{Preliminaries}

First, we introduce some basic concepts and notations which are
frequently used in this paper.

\begin{itemize}
\item[-] For a given prime $p$,  let $\theta$ and $\eta$ denote the
$(p-1)$th and $p$th primitive roots of unity in complex field
respectively, i.e.,
$$\theta=\exp \left( \frac{2\pi i}{p-1}\right)\ \  \mbox{and} \ \ \eta=\exp\left(\frac{2\pi i}{p}\right).$$
\item[-] We denote $\mathbb{F}_p$ as the finite field with $p$ elements,
and $\mathbb{F}_p^*=\mathbb{F}_p\backslash \{0\}$ as the
multiplicative group of $\mathbb{F}_p$ with a generator $a$. Then
for every element $b\in \mathbb{F}_p^*$, there exist $i$ with
$0\leqslant i\leqslant p-2$, such that $b=a^i$. In other words,
$i=\log_a b$.

\item[-] $SL_2(\mathbb{F}_p)$ is the 2-dimensional special linear group
over $\mathbb{F}_p$ consisting of all the $2\times 2$ matrices
$\left(
                                                      \begin{array}{cc}
                                                        a & b \\
                                                        c & d \\
                                                      \end{array}
                                                    \right)$
such that $ad-bc=1$ where $a,b,c,d \in \F_p$.

\item[-] Let
$\mathcal{H}=\C(\F_p)$ which is the $p$-dimensional Hilbert space
containing all the function from $\F_p$ to $\C$ with the standard
inner product, and $U(\mathcal{H})$ be the group of unitary
operators on $\mathcal{H}$.

\item[-]A unitary representation of a group $G$ on the Hilbert space
$\mathcal{H}$ is a homomorphism $\rho: G \rightarrow U(\mathcal{H})$
which satisfies $\rho(g\cdot h)=\rho(g)\cdot \rho(h)$ for $\forall
g,h \in G$. Specially, if $G$ is an Abelian group, its
representation $\rho$ can be decomposed into a direct sum of
1-dimensional representation (character).

\end{itemize}

\subsection{The Heisenberg Representation}

Let $(V ,\omega)$ be a two-dimensional symplectic vector space over
the finite field $\mathbb{F}_p$. For $ \forall  (t_i,w_i)\in
V=\mathbb{F}_p\times\mathbb{F}_p$ ($i=1,2$), the symplectic form
$\omega$ is given by
$$\omega((t_1,w_1),(t_2,w_2))=t_1w_2-t_2w_1.$$
Considering $V$ as an Abelian group, it admits a non-trivial central
extension called the {\em Heisenberg group} $H$. The group $H$ can
be presented as $H = V\times \F_p$ with the multiplication given by
$$(t_1,w_1,z_1)\cdot(t_2,w_2,z_2)=(t_1+t_2,w_1+w_2,z_1+z_2+2^{-1}\omega((t_1,w_1),(t_2,w_2))).$$
It is easy to verify the center of $H$ is $Z=Z(H)=\{g\in H: gH=Hg\}
= \{(0,0, z) : z\in \mathbb{F}_p$\}.

For a given non-trivial one dimensional representation $\phi$ of the
center $Z$, the Heisenberg group $H$ admits a unique irreducible
representation of $H$.

\begin{theorem}
(Stone-Von Neuman) Up to isomorphism, there exists a unique
irreducible unitary representation $\pi : H \rightarrow
U(\mathcal{H})$ with central character $\phi$, that is,
$\pi_{|Z}=\phi\cdot Id_{\mathcal{H}}$.
\end{theorem}

The representation $\pi$ which appears in the above theorem is
called the {\em Heisenberg representation}. In this paper, we take
one dimensional representation of $Z$ as $\phi((0,0,z))=\eta^z$.
Then the unique irreducible unitary representation $\pi$
corresponding to $\phi$ has the following formula
\begin{equation}
\pi(t,w,z)[\varphi](i)=\eta^{2^{-1}tw+z+wi}\varphi(i+t)
\end{equation}
for $\forall \varphi\in \mathcal{H}$, $(t,w,z)\in H$.

\subsection{The Weil Representation}
The symplectic group $Sp = Sp(V ,\omega)$, which is isomorphic to
$SL_2(\mathbb{F}_p)$, acts by automorphism of $H$ through its action
on the $V$-coordinate, i.e., for $\forall(t,w,z)\in H$ and a matrix
$g=\left(
     \begin{array}{cc}
       a & b \\
       c & d \\
     \end{array}
   \right)
\in SL_2(\mathbb{F}_p)$, the action $g$ on $(t,w,z)$ is defined as
\begin{equation}
g\cdot (t,w,z)= (at+bw, ct+dw, z).
\end{equation}

Let $GL(\mathcal{H})$ be the $p-$dimensional general linear group
over $\C$, and $PGL(\mathcal{H})$ be the projective general linear
group where $PGL(\mathcal{H})=GL(\mathcal{H})/\C^*$. Due to Weil
\cite{A.Weil}, a projective unitary representation
$\widetilde{\rho}: SL_2(\mathbb{F}_p)\rightarrow PGL(\mathcal{H})$
is constructed as follows. Considering the Heisenberg representation
$\pi: H \rightarrow U(\mathcal{H})$ and $\forall g\in
SL_2(\mathbb{F}_p)$, a new representation is define as: $\pi^g: H
\rightarrow U(\mathcal{H})$ by $\pi^g(h) = \pi(g(h))$. Because both
$\pi$ and $\pi^g$ have the same central character $\phi$, they are
isomorphic by Theorem $1$. By Schur's Lemma \cite{J. P. Serre},
$Hom_H$($\pi$,$\pi^g$)$\cong \mathbb{C}^*$, so there exist a
projective representation $\widetilde{\rho}:
SL_2(\mathbb{F}_p)\rightarrow PGL(\mathcal{H})$. This projective
representation $\widetilde{\rho}$ is characterized by the formula:
\begin{equation}\label{eq-we}
\widetilde{\rho}(g)\pi(h)\widetilde{\rho}(g^{-1})=\pi(g(h))
\end{equation}
for every $g\in SL_2(\mathbb{F}_p)$ and $h\in H$. A more delicate
statement is that there exists a unique lifting of
$\widetilde{\rho}$ into a unitary representation.

\begin{theorem}
The projective Weil representation uniquely lifts to a unitary
representation
$$\rho: SL_2(\mathbb{F}_p)\rightarrow U(\mathcal{H})$$
that satisfies equation (\ref{eq-we}).
\end{theorem}

The existence of $\rho$ follows from the fact \cite{Beyl} that any
projective representation of $SL_2(\mathbb{F}_p)$ can be lifted to
an honest representation, while the uniqueness of $\rho$ follows
from the fact \cite{S.Gurevich4} that the group $SL_2(\mathbb{F}_p)$
has no non-trivial characters when $p\neq 3$.

Note that $SL_2(\mathbb{F}_p)$ can be generated by $g_a=\left(
\begin{array}{cc}
a & 0 \\
0 & a^{-1} \\
\end{array}
\right)$, $g_b=\left(
\begin{array}{cc}
 1 & 0 \\
 b & 1 \\
 \end{array}
 \right)$, and
$w=\left(
\begin{array}{cc}
 0 & 1 \\
  -1 & 0 \\
  \end{array}
 \right)$
where $a\in \mathbb{F}_p^*$, $b \in \mathbb{F}_p$. The formulae of
their respective Weil representations for $g_a,g_b$ and $w$ are
given in \cite{S.Gurevich2} as follows
\begin{equation}\label{eq-sc1}
\rho(g_a)[\varphi](i)=\sigma(a)\varphi(a^{-1}i)\\
\end{equation}
\begin{equation}\label{eq-ch1}
\rho(g_b)[\varphi](i)=\eta^{-2^{-1}bi^2}\varphi(i)
\end{equation}
\begin{equation}\label{eq-fo1}
\rho(w)[\varphi](j)=\frac{1}{\sqrt{p}}\sum_{i\in
\mathbb{F}_p}\eta^{ji}\varphi(i)
\end{equation}
where $\sigma: \mathbb{F}_p^*\rightarrow\{\pm 1\}$ is the {\em
Legendre character}, i.e., $\sigma (a)=a^{\frac{p-1}{2}}$ in
$\mathbb{F}_p$.

Obviously, $\rho(w)=F$ which is the DFT, and we denote
$\rho(g_a)=S_a, \rho(g_b)=N_b$ for convenience. For $\forall g
=\left(
\begin{array}{cc}
a & b \\
c & d \\
\end{array}
\right) \in SL_2(\mathbb{F}_p)$, if $b\neq 0$,
$$g=\left(
\begin{array}{cc}
 a & b \\
 c & d \\
 \end{array}
 \right)=
 \left(
\begin{array}{cc}
a & b \\
(ad-1)b^{-1} & d \\
\end{array}
\right)= \left(
\begin{array}{cc}
b & 0 \\
0 & b^{-1} \\
\end{array}
\right) \left(
\begin{array}{cc}
1 & 0 \\
bd & 1 \\
\end{array}
\right) \left(
\begin{array}{cc}
0 & 1 \\
-1 & 0 \\
\end{array}
\right) \left(
\begin{array}{cc}
1 & 0 \\
ab^{-1} & 1 \\
\end{array}
\right).$$ Then the Weil representation of $g$ is given by
\begin{equation}
\rho(g)=S_{b}\circ N_{bd}\circ F\circ N_{ab^{-1}}.
\end{equation}
If $b=0$, then
$$g=\left(
\begin{array}{cc}
 a & b \\
 c & d \\
 \end{array}
 \right)=\left(
\begin{array}{cc}
 a & 0 \\
 c & a^{-1} \\
 \end{array}
 \right)=
 \left(
\begin{array}{cc}
 a & 0 \\
 0 & a^{-1} \\
 \end{array}
 \right)\left(
\begin{array}{cc}
 1 & 0 \\
 ac & 1 \\
 \end{array}
 \right).$$
Hence the Weil representation of $g$ is as follows
\begin{equation}
\rho(g)=S_a\circ N_{ac}.
\end{equation}

For more details about the Heisenberg and Weil representations,
please see \cite{S.Gurevich} \cite{S.Gurevich2} \cite{S.D.Howard}
\cite{R.Howe}.

\subsection{Centralizer Subgroup of the DFT}

\vspace{0.1in} \noindent
{\bf A. Maximal Algebraic Tori and $T_w$}

A maximal algebraic {\em torus} \cite{Borel} in $SL_2(\mathbb{F}_p)$
is a maximal commutative subgroup which becomes diagonalizable over
the original field or quadratic extension of the field. There are
two classes of tori in $SL_2(\mathbb{F}_p)$. The first class, called
{\em split tori}, consists of those tori which are diagonalizable
over $\mathbb{F}_p$, while the second class, called {\em non-split
tori}, consists of those tori which are not diagonalizable over
$\mathbb{F}_p$, but become diagonalizable over the quadratic
extension $\mathbb{F}_{p^2}$.

$T_w=\{g: gw=wg, g\in SL_2(\mathbb{F}_p)\}$ is the centralizer group
of $w$ in $SL_2(\mathbb{F}_p)$. It is easy to verify
\begin{equation}
T_w=\left\{\left(
 \begin{array}{cc}
  a & -b \\
  b & a \\
  \end{array}
  \right): a^2+b^2=1, a,b \in \mathbb{F}_p \right\}.
\end{equation}
If $p\equiv 1$ (mod 4), then $T_w$ is a split torus and conjugates
to the standard diagonal torus
$$A=\left\{\left(
\begin{array}{cc}
 a & 0 \\
 0 & a^{-1} \\
 \end{array}
 \right): a\in \mathbb{F}_p^*
\right\}.$$ So $T_w$ is a cyclic subgroup of $SL_2(\mathbb{F}_p)$
with order $p-1$. If $p\equiv 3$ (mod 4), $T_w$ is a non-split torus
which is a cyclic subgroup of $SL_2(\mathbb{F}_p)$ with order $p+1$.

\vspace{0.1in} \noindent {\bf B. Decomposition of Weil
representation Associated with $T_w$}

Because $T_w$ is a cyclic group, restricting the Weil representation
to $T_w$: $\rho_{|T_w}: T_w\rightarrow U(\mathcal{H})$, we obtain a
1-dimensional subrepresentation decomposition of $\rho_{|T_w}$
corresponding to an orthogonal decomposition of $\mathcal{H}$(see
\cite{J. P. Serre} for basics of group representation theory). In
other words,
\begin{equation}\label{eq-DE}
\rho_{|{T_w}}=\bigoplus_{\chi\in \Lambda_{T_w}}\chi\ \ \ \
\mbox{and} \ \ \
\mathcal{H}=\bigoplus_{\chi\in\Lambda_{T_w}}\mathcal{H}_\chi
\end{equation}
where $\Lambda_{T_w}$ is a collection of all the 1-dimensional
subrepresentation (character) $\chi: T_w\rightarrow \mathbb{C}$ in
the decomposition of Weil representation restricted to $T_w$.

If $p\equiv 1$ (mod 4), $\chi$ is the character given by $\chi:
\mathbb{Z}_{p-1}\rightarrow \mathbb{C}$. We have $dim
\mathcal{H}_{\chi} = 1$ unless $\chi=\sigma$ where $\sigma$ is the
Legendre character of $T$, and $dim \mathcal{H}_{\sigma} = 2$. If
$p\equiv 3$ (mod 4), $\chi$ is the character given by $\chi:
\mathbb{Z}_{p+1}\rightarrow \mathbb{C}$. There is only one character
which does not appear in the decomposition. For the other $p$
characters $\chi$ which appear in the decomposition, we have $dim
\mathcal{H}_{\chi} = 1$.

Choosing a generator $t\in T_w$, the character is generated by the
eigenvalue $\chi(t)$ of the linear operator $\rho(t)$, and the
character space $\mathcal{H}_\chi$ naturally corresponds to the
eigenspace of $\chi(t)$. Because the eigenvalues of $\rho(t)$ are
almost different, it is easier to find a basis of orthogonal
eigenvectors of $\rho(t)$ than of the DFT. Since $\rho(t)$ commutes
with the DFT, the eigenvectors of $\rho(t)$ are also the
eigenvectors of the DFT. Thus, we obtain a canonical basis of the
DFT.

\section{A Canonical Basis of the DFT}
If $p\equiv 1$ (mod 4), $T_w$ is a torus conjugating to the standard
diagonal torus $A=\left\{\left(
\begin{array}{cc}
 a & 0 \\
 0 & a^{-1} \\
 \end{array}
 \right): a\in \mathbb{F}_p^*
\right\},$ so $t$, which is a generator of $T_w$, conjugates to
$g_a=\left(
\begin{array}{cc}
 a & 0 \\
 0 & a^{-1} \\
 \end{array}
 \right)$ where $a$ is a generator of $\mathbb{F}_p^*$, i.e., there
exist $s \in SL_2(\mathbb{F}_p)$, such that $t=s g_a s^{-1}$ and
$\rho(t)=\rho(s)\rho(g_a)\rho(s^{-1}).$ Thus, the eigenvectors of
$\rho(t)$ can be determined by $\rho(s)$ and the eigenvectors of
$\rho(g_a)$. In the following, we first present the results, then
their proofs follow.

\begin{lemma}
Let $g_a=\left(
\begin{array}{cc}
 a & 0 \\
 0 & a^{-1} \\
 \end{array}
 \right)$ where $a$ is a generator of $\mathbb{F}_p^*$, then $\{\psi_x=\{\psi_x(i)\}_{0\leqslant i< p}: 0\leqslant x<p\}$ where $$\psi_0(i)=\left\{
\begin{aligned}
1&, \ for\  i=0 \\
0&, \ for\  i\neq 0
\end{aligned} \right.  \ \ and\ \
\psi_x(i)=\left\{ \begin{aligned}
0,\ \ \ \ \ \ \ \  \ \ \ \ \ \ \   &\ for\  i=0 \\
\frac{1}{\sqrt{p-1}}\theta^{x\log_ai}, &\ for\  i\neq 0
\end{aligned} \right.  \ \ for \ \ 0<x<p$$
is an orthonormal basis of $\mathcal{H}$ and a collection of the
eigenvectors of $\rho(g_a)$.
\end{lemma}

\begin{lemma}
Let $s=\left(
     \begin{array}{cc}
       1 & 2^{-1}a^k \\
       a^k & 2^{-1} \\
     \end{array}
   \right)$ where $k=\frac{p-1}{4}$, then $t=sg_as^{-1}$ is a generator of
   $T_w$.
\end{lemma}

Thus $\Phi_p=\{\varphi_x: \varphi_x=\rho(s)\psi_x, 0\leqslant x<p\}$
is a canonical basis of $\rho(t)=\rho(sg_as^{-1})$ and the DFT. More
explicitly,

\begin{theorem}
Let
$$\varphi_x(i)=\left\{ \begin{aligned}
\frac{1}{\sqrt{p}}\eta^{2^{-1}a^ki^2},\ \ \ \ \ \ \ \ \ \ \ \ \ \ \ \ \ \ \ \ \  \ \ \ \ \ \ \ \ \ \ &\ for\  x=0 \\
\frac{1}{\sqrt{p(p-1)}}\sum_{j=1}^{p-1}\theta^{x\log_aj}\eta^{a^k(j-i)^2-2^{-1}a^ki^2},
&\ for\ 0<x<p.
\end{aligned} \right. $$
Then $\Phi_p=\{\varphi_x=\{\varphi_x(i)\}_{0\leqslant i<p}:
0\leqslant x<p\}$ is an orthonormal basis of $\mathcal{H}$ and a
collection of the eigenvectors of the DFT.
\end{theorem}

\begin{theorem}
$\varphi_x (0\leqslant x<p)$ is the eigenvector of the DFT
corresponding to the eigenvalue $(-i)^x$ where $i=\sqrt{-1}$, i.e.,
$$F\varphi_x=(-i)^{x}\varphi_x.$$
\end{theorem}

Now we prove the above lemmas and theorems. Considering $\{\delta_i:
i\in \mathbb{F}_p\}$ which is the orthonormal basis of Hilbert space
$\mathcal{H}=\mathbb{C}(\mathbb{F}_p)$, where $\delta_i$ is defined
as $\delta_i(j)=\delta_{ij}$ for $\forall i,j\in\mathbb{F}_p$, every
vector $\varphi=\{\varphi(i)\}$ can be written as the form
$\varphi=\sum_{i\in \F_p}\varphi(i)\delta_i$. Recall that
$SL_2({\mathbb{F}_p})$ can be generated by  $g_a=\left(
\begin{array}{cc}
a & 0 \\
0 & a^{-1} \\
\end{array}
\right)$, $g_b=\left(
\begin{array}{cc}
 1 & 0 \\
 b & 1 \\
 \end{array}
 \right)$ and
$w=\left(
\begin{array}{cc}
 0 & 1 \\
  -1 & 0 \\
  \end{array}
 \right)$
where $a\in \mathbb{F}_p^*$ and $b \in \mathbb{F}_p$. Then their
respective Weil representations (\ref{eq-sc1}),(\ref{eq-ch1}), and
(\ref{eq-fo1}) of $g_a$, $g_b$, and $w$ can be rewritten as follows
\begin{equation}\label{eq-sc2}
\rho(g_a)\delta_i=S_a\delta_i=\sigma(a)\delta_{ai}\\
\end{equation}
\begin{equation}\label{eq-ch2}
\rho(g_b)\delta_i=N_b\delta_i=\eta^{-2^{-1}bi^2}\delta_i
\end{equation}
\begin{equation}\label{eq-fo2}
\rho(w)\delta_j=F\delta_j=\frac{1}{\sqrt{p}}\sum_{i\in
\mathbb{F}_p}\eta^{ji}\delta_i.
\end{equation}

\vspace{0.1in} \noindent {\bf Proof of Lemma 1.} From
(\ref{eq-sc2}), we have
$$\rho(g_a)\delta_i=\sigma(a)\delta_{ai}=-\delta_{ai}.$$
Let $V_1=V(\delta_1), V_2=V(\delta_2, \delta_3, \cdots,
\delta_{p-1})$, then it is obvious that $\mathcal{H}=V_1\bigoplus
V_2$, $<V_1, V_2>=0$, and $\rho(g_a)(V_i)=V_i$ for $i=1,2$. It is
easy to see that $\rho(g_a)_{|V_1}=-Id$, so $\delta_0$ is a
eigenvector of $\rho(g_a)$ corresponding to the eigenvalue $-1$. The
eigenfunction of $\rho(g_a)_{|V_2}$ is $(x^{p-1}-1$), so the
eigenvalues of $\rho(g_a)_{|V_2}$ are $\theta^0,\theta^1, \theta^2,
\cdots,  \theta^{p-2}$ which are different. We assert that
$\sum_{i=1}^{p-1}\theta^{(\frac{p-1}{2}-j)\log_a i}\delta_i$ is the
eigenvector associated to the eigenvalue $\theta^j (0\leqslant
j\leqslant p-2)$, and it can be verified as follows
\begin{eqnarray*}
\rho(g_a)(\sum_{i=1}^{p-1}\theta^{(\frac{p-1}{2}-j)\log_a
i}\delta_i) &=&-\sum_{i=1}^{p-1}\theta^{(\frac{p-1}{2}-j)\log_a
i}\delta_{ai}\\
&=&-\sum_{i=1}^{p-1}\theta^{(\frac{p-1}{2}-j)\log_a (a^{-1} i)}\delta_i\\
&=&\theta^{\frac{p-1}{2}}\sum_{i=1}^{p-1}\theta^{(\frac{p-1}{2}-j)(\log_a i-1)}\delta_i\\
&=&\theta^{\frac{p-1}{2}}\theta^{j-\frac{p-1}{2}}\sum_{i=1}^{p-1}\theta^{(\frac{p-1}{2}-j)\log_a
i}\delta_i\\
&=&\theta^j\sum_{i=1}^{p-1}\theta^{(\frac{p-1}{2}-j)\log_a
i}\delta_i.
\end{eqnarray*}
Let $x=\frac{q-1}{2}-j$. By normalizing the eigenvectors, we
complete the proof. \done

\vspace{0.1in} \noindent {\bf Proof of Lemma 2.} Note that
\begin{eqnarray*}
t&=&sg_as^{-1}\\
&=&\left(
     \begin{array}{cc}
       1 & 2^{-1}a^k \\
       a^k & 2^{-1} \\
     \end{array}
   \right)\left(
            \begin{array}{cc}
              a & 0 \\
              0 & a^{-1}\\
            \end{array}
          \right)
          \left(
     \begin{array}{cc}
       2^{-1} & -2^{-1}a^k \\
       -a^k & 1 \\
     \end{array}
   \right)
   \\
&=& \left(
     \begin{array}{cc}
       2^{-1}(a-a^{-1}) & -2^{-1}a^k(a-a^{-1}) \\
       2^{-1}a^k(a-a^{-1}) & 2^{-1}(a-a^{-1}) \\
     \end{array}
   \right)\in T_w.
\end{eqnarray*}
On the other hand, $t$ conjugates to $g_a$, so the order of $t$ is
$p-1$. Thus, $t$ is a generator of $T_w$.

\vspace{0.1in} \noindent {\bf Proof of Theorem 3.} Since
$t=sg_as^{-1}$, $\Phi_p=\{\varphi_x=\rho(s)\psi_x: 0\leqslant x<p\}$
is a collection of the orthogonal eigenvectors of $\rho(t)$ where
$\varphi_x$ and $s$ are presented in Lemmas 1 and 2 respectively.

From (\ref{eq-ch2}), $s$ has the following decomposition
$$s=\left(
  \begin{array}{cc}
    1 & 2^{-1}a^k \\
    a^k & 2^{-1} \\
  \end{array}
\right)= \left(
\begin{array}{cc}
2^{-1}a^k & 0 \\
0 & 2a^{3k} \\
\end{array}
\right) \left(
\begin{array}{cc}
1 & 0 \\
4^{-1}a^k & 1 \\
\end{array}
\right) \left(
\begin{array}{cc}
0 & 1 \\
-1 & 0 \\
\end{array}
\right) \left(
\begin{array}{cc}
1 & 0 \\
2a^{3k} & 1 \\
\end{array}
\right).$$ Then applying (\ref{eq-sc2}),(\ref{eq-ch2}), and
(\ref{eq-fo2}), for $1\leqslant x\leqslant p-1$, we have
\begin{eqnarray*}
\varphi_x=\rho(s)\psi_x&=&S_{2^{-1}a^k}\circ N_{4^{-1}a^k}\circ
F\circ
N_{2a^{3k}}(\frac{1}{\sqrt{p-1}}\sum_{j=1}^{p-1}\theta^{x\cdot
\log_a
j}\delta_j)\\
&=&S_{2^{-1}a^k}\circ N_{4^{-1}a^k}\circ
F(\frac{1}{\sqrt{p-1}}\sum_{j=1}^{p-1}\theta^{x\cdot \log_a
j}\eta^{a^kj^2}\delta_j)\\
&=&S_{2^{-1}a^k}\circ
N_{4^{-1}a^k}(\frac{1}{\sqrt{p(p-1)}}\sum_{i=0}^p\sum_{j=1}^{p-1}\theta^{x\cdot
\log_a j}\eta^{a^k j^2+ij}\delta_i)\\
&=&S_{2^{-1}a^k}(\frac{1}{\sqrt{p(p-1)}}\sum_{i=0}^p\sum_{j=1}^{p-1}\theta^{x\cdot
\log_a j}\eta^{a^k j^2+ij-8^{-1}a^ki^2}\delta_i)\\
&=&\frac{\sigma(2^{-1}a^k)}{\sqrt{p(p-1)}}\sum_{i=0}^p\sum_{j=1}^{p-1}\theta^{x\cdot
\log_a j}\eta^{a^k j^2+ij-8^{-1}a^ki^2}\delta_{2^{-1}a^ki}) \ \ (\mbox{substitute} \ \ i \ \ \mbox{by}\ \  2^{-1}a^ki)\\
&=&\frac{\sigma(2^{-1}a^k)}{\sqrt{p(p-1)}}\sum_{i=0}^p\sum_{j=1}^{p-1}\theta^{x\cdot
\log_a
j}\eta^{a^kj^2+2a^{-k}ij+2^{-1}a^ki^2}\delta_i\\
&=&\frac{\sigma(2^{-1}a^k)}{\sqrt{p(p-1)}}\sum_{i=0}^p\sum_{j=1}^{p-1}\theta^{x\cdot
\log_a
j}\eta^{a^k(j-i)^2-2^{-1}a^ki^2}\delta_i.\\
\end{eqnarray*}
For $x=0$, we have
$$\varphi_0=\rho(s)\phi_0=\frac{\sigma(2^{-1}a^k)}{\sqrt{p}}\sum_{i=0}^p\eta^{2^{-1}a^ki^2}\delta_i.$$

Because $\{\psi_x: 0\leqslant x<p \}$ is an orthonormal basis and
$\rho(s)$ is a unitary matrix, $\{\varphi_x: 0\leqslant x<p \}$ is
also an orthonormal basis of $\mathcal{H}$. Since
$\rho(s)F=F\rho(s)$, $\varphi_x (0\leqslant x<p)$ are not only the
eigenvectors of $\rho(s)$, but also the eigenvectors of the DFT.
Note that $\sigma(2^{-1}a^k)$ is a constant, which completes the
proof.\done

\vspace{0.1in} \noindent {\bf Proof of Theorem 4.} It can be
 verified as follows, for $x=0$, we have
\begin{eqnarray*}
F[\varphi_0](t)&=&\frac{1}{p}\sum_{t=0}^{p-1}\eta^{2^{-1}a^ki^2+it}\\
&=&\frac{1}{p}\eta^{2^{-1}a^kt^2}\sum_{t=0}^{p-1}\eta^{2^{-1}a^ki^2+it-2^{-1}a^kt^2}\\
&=&\frac{1}{p}\eta^{2^{-1}a^kt^2}\sum_{t=0}^{p-1}(\eta^{2^{-1}a^k})^{(i-a^kt)^2}\\
&=&\frac{1}{p}\eta^{2^{-1}a^kt^2}\sum_{t=0}^{p-1}(\eta^{2^{-1}a^k})^{t^2}\ \ \ \ (\mbox{substitute} \ \  i-a^kt \ \ \mbox{by}\ \  t)\\
&=&\frac{1}{\sqrt{p}}\eta^{2^{-1}a^kt^2}\ \ \ \ \ \ \ \ \ \ \ \ \ \ \ \ \ \ (\mbox{Gauss sum})\\
&=&\varphi_0(t).
\end{eqnarray*}
For $x \neq 0$, we have
\begin{eqnarray*}
F[\varphi_x](t)&=&\frac{1}{p\sqrt{p-1}}\sum_{t=0}^{p-1}\sum_{j=1}^{p-1}\theta^{x\log_aj}\eta^{a^k(j-i)^2-2^{-1}a^ki^2+it}\\
&=&\frac{1}{p\sqrt{p-1}}\sum_{j=1}^{p-1}\theta^{x\log_aj}\sum_{t=0}^{p-1}\eta^{a^k(j-i)^2-2^{-1}a^ki^2+it}\\
&=&\frac{1}{p\sqrt{p-1}}\sum_{j=1}^{p-1}\theta^{x\log_aa^{-k}j}\sum_{t=0}^{p-1}\eta^{a^k(a^{-k}j-i)^2-2^{-1}a^ki^2+it}\ \ \ \ (\mbox{substitute} \ \ j \ \ \mbox{by}\ \ a^{-k}j )\\
&=&\frac{\theta^{-kx}}{p\sqrt{p-1}}\sum_{j=1}^{p-1}\theta^{x\log_aj}\sum_{t=0}^{p-1}\eta^{a^{-k}j^2-2ji+2^{-1}a^ki^2+it}\\
&=&\frac{(-i)^{x}}{p\sqrt{p-1}}\sum_{j=1}^{p-1}\theta^{x\log_aj}\eta^{a^{-k}j^2-2^{-1}a^{-k}(t-2j)^2}\sum_{t=0}^{p-1}\eta^{2^{-1}a^ki^2+i(t-2j)+2^{-1}a^{-k}(t-2j)^2}\\
&=&\frac{(-i)^{x}}{p\sqrt{p-1}}\sum_{j=1}^{p-1}\theta^{x\log_aj}\eta^{a^k(j-t)^2-2^{-1}a^kt^2}\sum_{t=0}^{p-1}\eta^{(2a^k)^{-1}(a^ki+t-2j)^2}\\
&=&\frac{(-i)^{x}}{\sqrt{p(p-1)}}\sum_{j=1}^{p-1}\theta^{x\log_aj}\eta^{a^k(j-t)^2-2^{-1}a^kt^2}\ \ \ \ \ \ \ \ \ \ \ \ \ \ \ \ \ \ (\mbox{Gauss sum})\\
&=&(-i)^x\varphi_x(t).
\end{eqnarray*}

\section*{Acknowledgment}
The authors would like to thank Grevich, Hadani and Sochen for their
help  during the course of conducting this work.


\begin{thebibliography}{1}

\bibitem{Atakishiyev} N.M. Atakishiyev and K.B. Wolf, Fractional
Fourier-Kravchuk transform, {\em J. Opt. Soc. Am.,} Vol. 14, No. 7,
1997, pp. 1467-1477.

\bibitem{Borel} A. Borel, {\em Linear Algebraic Groups}. Graduate Texts in Mathematics,
vol. 126, Springer, New York, 1991.

\bibitem{Beyl} F.R. Beyl, The Schur multiplicator of $SL(2, \mathbb{Z}/m\mathbb{Z})$ and the congruence
subgroup property, {\em Math. Zeit}, 191, 1986.

\bibitem{Candan} C. Candan, M. A. Kutay and H. M.Ozaktas, The discrete
fractional Fourier transform, {\em IEEE Trans. Signal Processing}
Vol. 48, No.5, 2000, pp. 1329-1337.

\bibitem{Dickinson} B.W. Dickinson and K. Steiglitz, Eigenvectors and
functions of the discrete Fourier transform, {\em IEEE Trans.
Acoustics Speech and Signal Proc.}, Vol. 30, No. 1, 1982, pp. 25-31.


\bibitem{S.Gurevich} S. Gurevich, R. Hadani, and  N. Sochen. The finite harmonic oscillator
and its applications to sequences, communication and radar. {\em
IEEE Trans. Inform. Theory}, Vol. 54, No. 9, September 2008, pp.
4239-4253.

\bibitem{S.Gurevich2} S. Gurevich, R. Hadani, and  N. Sochen, On some deterministic
dictionaries supporting sparsity, {\em Journal of Fourier Analysis
and Applications}, Vol. 14, No. 5-6, December 2008, pp. 859-876.

\bibitem{S.Gurevich3}S. Gurevich, R. Hadani, and  N. Sochen, Group representation design of digital signals and
sequences,
 {\em the Proceedings of the International Conference on Sequences and Their Applications (SETA)}, 2008, Sep. 14-18, 2008, Lexington,
 KY, USA. {\em Sequences and Their Applications-SETA 2008}, LNCS 5203, S.W. Golomb, et al. (Eds.), Springer, 2008, pp. 153-166.

\bibitem{S.Gurevich4} S. Gurevich  and R. Hadani, On the diagonalization of the discrete Fourier
transform, {\em Applied and Computational Harmonic Analysis}, to
appear 2009.

\bibitem{Hanna} M.T. Hanna, N.P.A. Seif, and W.A.E.M. Ahmed, Hermite-Gaussian-like
eigenvectors of the discrete Fourier transform matrix based on the
singular-value decomposition of its orthogonal projection matrices,
{\em IEEE Trans. Circ. Syst. I}, Vol. 51, No. 11, 2004, pp.
2245-2254.



\bibitem{S.D.Howard}S.D. Howard, A.R. Calderbank, and W. Moran, The finite Heisenberg-Weyl groups in radar and communications, {\em EURASIP J. Appl.
Signal Process}, 2006, pp.1-12.


\bibitem{R.Howe} R. Howe,  Nice error bases, mutually unbiased bases, induced
representations, the Heisenberg group and finite geometries, {\em
Indag. Math. (N.S.)}, Vol. 16, No. 3-4, 2005, pp. 553-583.

\bibitem{McClellan} J. H. McClellan and T. W. Parks, Eigenvalues and
eigenvectors of the discrete Fourier transformation {\em IEEE Trans.
Audio Electroacoust,} Vol. 20, No. 1, 1972, pp. 66-74.


\bibitem{J. P. Serre} J.P. Serre, {\em Linear Representations of Finite Groups}. Graduate Texts
in Mathematics, Vol. 42, Springer, New York, 1977.




\bibitem{B. L. van der Waerden} B. L. van der Waerden, {\em Moderne
Algebra}, Springer, 1931.

\bibitem{Wang} Z. Wang, G. Gong, New sequences design from Weil representation with low two-dimensional correlation in both time and phase
shifts, http://arxiv.org/abs/0812.4487, Technical Report 2009-1,
University of Waterloo, 2009.


\bibitem{A.Weil} A. Weil, Sur certains groupes d'operateurs unitaires, {\em Acta.
Math.}, Vol. 111, 1964, pp. 143-211.



\end{thebibliography}
\end{document}